\documentclass[prb,twocolumn,amsmath,amssymb,showpacs]{revtex4}
\usepackage{graphicx}

\def\beq{\begin{eqnarray}}
\def\eeq{\end{eqnarray}}
\def\beqa{\begin{eqnarray}}
\def\eeqa{\end{eqnarray}}

\begin{document}

\title{$t-J$ model one-electron renormalizations: high energy features in 
photoemission experiments of high-$T_c$ cuprates.
}

\author{Andr\'es Greco}
\affiliation{
Facultad de Ciencias Exactas, Ingenier\'{\i}a y Agrimensura and
Instituto de F\'{\i}sica Rosario
(UNR-CONICET).
Av. Pellegrini 250-2000 Rosario-Argentina.
}

\date{\today}

\begin{abstract}
Recent angle-resolved photoemission experiments in hole doped cuprates 
reported new and 
interesting 
high energy features which may be useful for understanding the electronic properties 
of these materials. 
Using a perturbative approach, which allows the calculation of dynamical properties
in the $t-J$ model, one-electron spectral properties 
were calculated.
A strongly renormalized quasiparticle band near the Fermi surface and incoherent 
spectra at high energy were obtained. 
Among different current experimental interpretations, the obtained results 
are closer to the interpretation  
given by Pan {\it et al.}\cite{pan}.
The self-energy shows  large high energy  contributions  
which are responsible for the incoherent structures showed by the spectral functions and 
the reduction of the quasiparticle weight and bandwidth. 
According to the calculation, collective charge fluctuations are the main source 
for the self-energy renormalizations.
For testing if the obtained self-energy is compatible with transport measurement the 
resistivity versus temperature was estimated.
\end{abstract}

\pacs{71.10.Fd, 71.27.+a, 79.60.-i}

\maketitle

The understanding of high-$T_c$ cuprates is one of the mayor challenges
in solid state physics. Even with the problem unresolved, it is clear 
that not only the large value of the superconducting $T_c$ is anomalous. 
Cuprates have also in common many electronic properties which are in clear contrast 
with the expected ones in usual metals. One of these, which is the subject of the 
present paper, is the one-electron renormalization obtained by 
angle-resolved photoemission spectroscopy ($ARPES$).
Some years ago $ARPES$ 
reported a kink in the electronic 
dispersion,  at about $\sim 50-70 meV$,
of hole doped cuprates 
\cite{valla99,bogdanov00,kaminski00}.
This kink indicates the presence of a small energy scale in the electronic self-energy. 
Besides the kink, early $ARPES$ experiments \cite{valla99} reported 
an imaginary part of the self-energy without sign
of saturation up to energies 
of the order of $150-200 meV$. This feature, which was recovered in further 
experiments (see for instance Ref.[\onlinecite{kordyuk05}]),  
indicates that besides 
low energy, high energy excitations are also present.
However, since $ARPES$ experiments were reported only for $\omega < 300 meV$ 
this discussion was postponed until very recently.

Recently, $ARPES$ measurements 
\cite{valla,xie,graf,chang,meevasana,pan} reported results up to large energy 
$\omega \sim -1eV$, clearly showing the presence of high energy  
self-energy renormalizations contributing to the spectral functions.
The extracted $E-{\bf k}$ dispersion from momentum distribution curves 
seems to show a nearly 
vertical ``dive''\cite{pan} (also called ``waterfall''\cite{graf}) at about $350 meV$. 
These experiments provide opportunity for new investigations about the  
electronic order behind cuprates.
In spite of different experiments showing similar features, 
the interpretation is not unique \cite{valla,xie,graf,chang,meevasana,pan}.
For instance, 
Xie {\it et al.} \cite{xie} argue that, near the Fermi level,  the 
quasiparticle band breaks 
at about $\sim -350 meV$ and, at higher energies follows 
the dispersion predicted by band structure calculation. 
Graf {\it et al.} \cite{graf} interpreted their results in terms of the disintegration
of low energy Zhang-Rice singlet and the re-emergency of the band structure dispersion 
at high energies. 
Pan {\it et al.} \cite{pan} state that, at low energy,  there is a 
strongly renormalized coherent band,  
while the spectra is incoherent at high energies. In this picture the vertical 
``dive'' can be seem as a coherent-incoherent crossover and therefore, 
high energy features are not related with band structure calculations.

\begin{figure}
\vspace{1cm}
\begin{center}
\setlength{\unitlength}{1cm}
\includegraphics[width=8cm,angle=0]{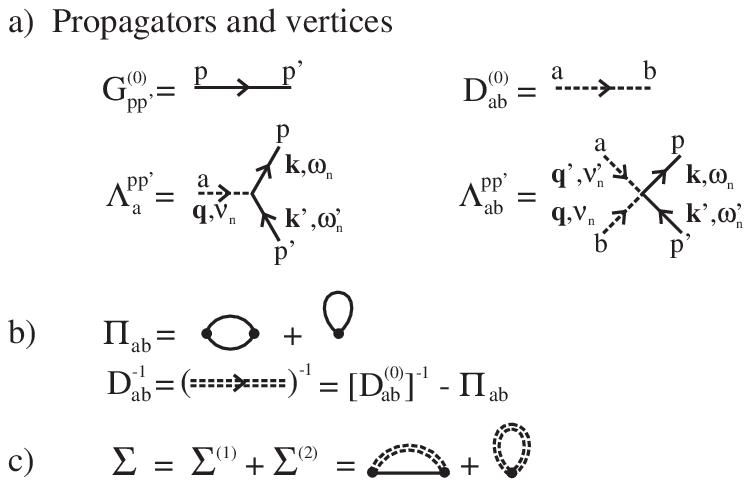}
\end{center}
\caption{
a) Solid line is the propagator, which is $O(1)$, for an electron  
with the dispersion $E_k$ described in the text.
Dashed line is the $6 \times 6$ boson propagator, which is $O(1/N)$, for the
six component boson field $\delta X^a$. The component $\delta X^1$ corresponds to 
charge fluctuations, $\delta X^2$ is introduced to 
fulfil the non-double occupancy constraint, and $\delta X^a$, with $a$ 
from $3$ to $6$, 
is associated with the Heisenberg coupling  $J$. 
$\Lambda^{pp'}_{a}$ and    
$\Lambda^{pp'}_{ab}$ are the interaction vertices between two fermions and one and two 
bosons respectively. Vertices are $O(1)$ and were obtained from the effective theory 
constructed under the requirement that non-double occupancy 
and the Hubbard operators algebra 
be satisfied. 
Combining the order of vertices and propagators a given physical quantity 
can be evaluated at a given order of $1/N$. This counting means that 
the approach is controllable by the small parameter $1/N$.
b) Irreducible boson self-energy $\Pi_{ab}$ and the renormalized boson 
propagator (double dashed line). 
c) Contributions $\Sigma^{(1)}$ and $\Sigma^{(2)}$ 
to the electron self-energy $\Sigma({\bf k},\omega)$ through 
$O(1/N)$. In $\Sigma$, double dashed line, which contains collective charge 
fluctuations,  
can be seen as the excitations that interacting with fermions 
lead to the self-energy effects and incoherent structures 
discussed in the text. 
}\label{FD}
\end{figure}

In this paper,  
electronic spectral functions and 
self-energy corrections are investigated in the framework of the $t-J$ model.
The obtained results are confronted with the experiments
suggesting support for the scenario proposed by Pan {\it et al.}. 
The calculation of spectral properties in the $t-J$ model requires 
a controllable treatment of the non-double-occupancy constraint.
While there are many calculations at mean field level, the  
evaluation of fluctuations above mean field, 
which is of interest for understanding dynamical properties as, for instance, 
the electronic 
self-energy, 
is very hard. 
Recently we have developed a large-$N$ perturbative approach\cite{foussats04}
(where the spin components have been generalized to $N$ components)  
for the $t-J$ model 
based on the path integral representation for Hubbard operators. 
The advantage of this approach rests on the fact that it is formulated in terms 
of Hubbard operators as fundamental objects and, 
since  
there is no any decoupling scheme, problems that arise in other 
treatments are avoided, like considering 
fluctuations of the gauge field or Bose condensation that appears in the slave 
boson approach\cite{lee06}.
It is not our aim to give here a detailed description of the method, it can be found
in Refs.[\onlinecite{foussats04,bejas06}],
only a brief 
summary   
is given in Fig.\ref{FD}. 
Using the Feynman diagrams (Fig.\ref{FD}a), the self-energy $\Sigma({\bf k},\omega)$ 
can be evaluated (Fig.\ref{FD}c) and
with it, the spectral function $A({\bf k},\omega)$ can be obtained as usual.
In Ref.[\onlinecite{bejas06}], in order to test  the confidence  
of our method, spectral functions were compared with those 
obtained using Lanczos diagonalization finding fairly good agreement.
Also high energy self-energy excitations were identified but not compared with the new $ARPES$ experiments\cite{valla,xie,graf,chang,meevasana,pan} which are more recent than Ref.[\onlinecite{bejas06}].

\begin{figure}
\vspace{1cm}
\begin{center}
\setlength{\unitlength}{1cm}
\includegraphics[width=8cm,angle=0]{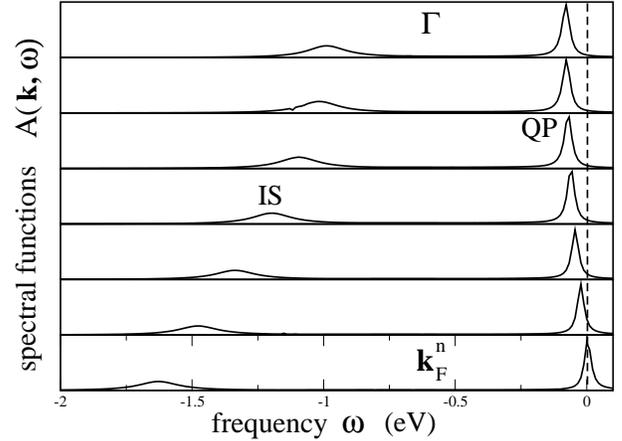}
\end{center}
\caption{
Spectral functions along the nodal direction from $\Gamma$ to the nodal 
Fermi vector ${\bf k}_F^n$. Close to the Fermi surface a strongly renormalized 
quasiparticle (QP) coherent band
is obtained. For large energy ($\sim -1.eV$) incoherent structure (IS) is 
observed. The vertical dashed line marks the Fermi level. 
All vertical scales are equal.
When ${\bf k}$ moves from
$\Gamma$ to ${\bf k}_F^n$, 
while the quasiparticle peak approaches $\omega=0$, 
the incoherent structure moves in opposite direction
in qualitative  
agreement with the experiment. See text for discussions. 
} \label{SF}
\end{figure}

Once presented the problem and the general characteristics of the method,
results  
for the $tt'-J$ model are given. 
$t$ and $t'$
are the nearest and second-nearest neighbor hopping amplitudes respectively, 
and $J$ is the 
Heisenberg coupling. In what follows we choose 
$t'/t=0.35$, $J/t=0.3$ \cite{dagotto} and the calculation was done in the normal state.
At mean field level the obtained electronic band is  
($E_k=-2(t \delta/2+\Delta)(cos(k_x)+cos(k_y))+4 t'\delta/2 cos (k_x) cos(k_y)-\mu$)
where ($\Delta=J/2N_s \sum_k cos (k_x) n_F(E_k)$) and $\mu$ the chemical potential.
$N_s$ is the number of sites and $n_F$ the Fermi function.
The bare (or mean field) band $E_k$ (which already at this level 
is renormalized by correlations as shown  
by the presence of the doping $\delta$ and $J$) will be dynamically dressed by 
$\Sigma({\bf k}, \omega)$. 
For these parameters,
in the doping range of interest for cuprates,
a hole-like Fermi surface is obtained.
We also choose $\delta=0.26$ which corresponds to highly overdoped regime
where several $ARPES$ experiments were performed\cite{xie,pan}. 
On the other hand, as discussed in Ref.[\onlinecite{bejas06}], our method is better for large than for low
doping.  The existence of anomalous features in highly overdoped samples is very interesting because the system is far from the antiferromagnetic phase and the pseudogap, if it is not zero, is very weak.
For $\delta=0.26$   
the nodal Fermi vector is $k_F^n=(0.39,0.39)\pi/a$. 

\begin{figure}
\vspace{1cm}
\begin{center}
\setlength{\unitlength}{1cm}
\includegraphics[width=9cm,angle=0]{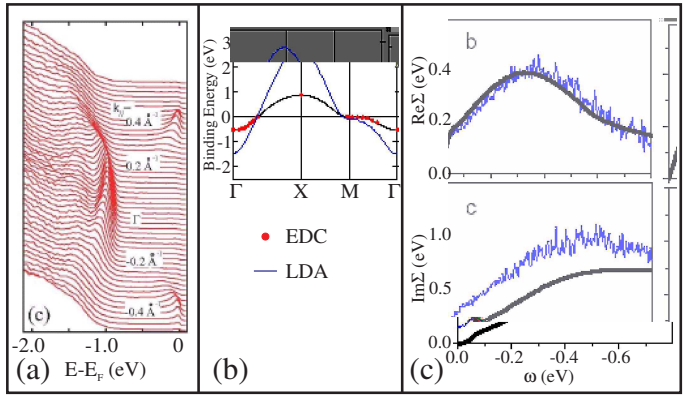}
\end{center}
\caption{
(a) Fig.1c modified from Ref.[\onlinecite{xie}] showing energy distribution curves 
(EDC) where low energy quasiparticle peaks 
and high energy features are observed. 
(b) Fig.4a modified from Ref.[\onlinecite{pan}] where a strongly renormalized 
coherent band (full circles) is reported near $\Gamma$. Notice the bandwidth 
reduction with respect to the band structure calculation (LDA).
(c) Figs.4b and 4c modified from Ref.[\onlinecite{valla}] showing the real and imaginary 
parts of the self-energy.
} \label{exp}
\end{figure}

\begin{figure}
\vspace{1cm}
\begin{center}
\setlength{\unitlength}{1cm}
\includegraphics[width=8cm,angle=0]{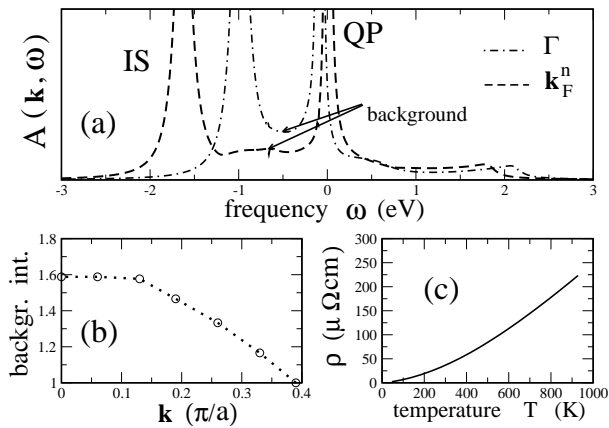}
\end{center}
\caption{
(a) Spectral functions for the $\Gamma$ and ${\bf k}_F^n$ vectors using  
an appropriate vertical scale in order to see the spectral weight 
in the background between 
the quasiparticle (QP) and the incoherent structure (IS). The spectral weight 
for $\omega >0$ is also shown.
(b) Background intensity (normalized to the background intensity at ${\bf k}_F^n$) vs 
${\bf k}$ from $\Gamma$ to ${\bf k}_F^n$. 
(a) and (b) show that the intensity of the background increases 
when approaching $\Gamma$.
(c)
Resistivity $\rho$ vs temperature $T$.  
$\rho$ was estimated using the procedure described in the text.
The resistivity values are in the order of magnitude 
of the experiments\cite{Ma,Timusk}.
In addition, $\rho$ vs $T$ presents a fractional power low $\rho \sim T^m$ where 
$m \sim 1.6-1.7$.
$\omega_p=2eV$ was chosen (see text).
} \label{SFp}
\end{figure}

Results for the spectral functions (energy distribution curves) 
along the nodal direction, from $\Gamma$ $(0,0)$ 
to 
${\bf k}_F^n$,  are presented in 
Fig.\ref{SF} where we adopt the accepted value $t=0.4eV$\cite{dagotto}.
Close to the Fermi surface a highly renormalized parabolic quasiparticle  
coherent band is obtained.
In addition, at high energy ($\sim -1eV$) incoherent structures are present.
For the present parameters the quasiparticle weight results 
$Z=(1-\frac{\partial Re \Sigma}{\partial \omega})^{-1} \sim 0.4$.
The remainder spectral weight lies mainly 
in the incoherent structure. There is also
spectral weight in the form of a tail  between the quasiparticle and the incoherent 
structure and at $\omega >0$ (Fig.\ref{SFp}a).

Let us compare Fig.\ref{SF} with 
the experiment
\cite{xie,pan}.
Similarly to Fig.1c in Xie {\it et al.}\cite{xie}, 
which is reproduced here in Fig.\ref{exp}a,  
Fig.\ref{SF} shows that 
while the low energy peak moves toward the Fermi surface, 
the high energy structure disperses in opposite direction.
A difference with the results in Xie {\it et al.}
is the following.
In their results the low energy peak is observed  
near ${\bf k}_F^n$, and away 
from it losses intensity being nearly invisible when approaching $\Gamma$ 
(Fig.\ref{exp}a).
This behavior is of fundamental interest for the ``waterfall'' interpretation. The
vanishing of the quasiparticle intensity near $\Gamma$ 
suggests that the low energy dispersion evolves abruptly to the high energy features.
It is important to notice that our Fig.\ref{SF}
does not exhibit  
the mentioned intensity decreasing away from the Fermi surface and shows  
well defined quasiparticles and incoherent structures for all {\bf k}-vectors.
It is not clear which is the reason for the experimental  
decreasing of the low energy peak intensity. Notice that  
the predicted quasiparticle weight
$Z$ is small, thus probably hard to follow experimentally away from the Fermi surface
because it may become mixed with the background as discussed below. 
In spite of the intensity decreasing  
away from 
${\bf k}_F^n$, 
Pan {\it et al.}\cite{pan} resolved  the quasiparticle peak approaching $\Gamma$ 
(see Fig.1e and Fig.4a in Ref.[\onlinecite{pan}], Fig.4a is reproduced here in 
Fig.\ref{exp}b) following
a parabolic shape and, at the same time,  high energy spectral features are observed 
(see Fig.1d in that paper) as in  our Fig.\ref{SF}. 
Our calculated quasiparticle bandwidth is somewhat smaller than in the experiment.
We think that the existence of low energy quasiparticle peaks near 
$\Gamma$ make doubtful the interpretation in terms of only one feature 
evolving from low  to high energies.  
Using Lanczos diagonalization on the $t-J$ model, similar  
high energy spectral features were reported in Ref.[\onlinecite{moreo}]
(see also Ref.[\onlinecite{stephan}]).

In Ref.[\onlinecite{pan}] it was observed that the ``diving'' 
behavior, which is mainly inferred from momentum distribution curves, is not manifested
in the energy distribution curves, instead, an enhancement of the background of the 
energy distribution curves 
is observed near $\Gamma$  
(see Fig.1f in Pan {\it et al.})  making difficult the quasiparticle peak detection. 
Thus, the background enhancement may be important for understanding 
differences between momentum and energy distribution curves. 
In Fig.\ref{SFp}a we present spectral function results at $\Gamma$ and 
${\bf k}_F^n$ using an appropriate  vertical scale in order to see 
the spectral weight in the background between the quasiparticle and the 
incoherent structure. In Fig.\ref{SFp}b the background intensity,
normalized to the background intensity at ${\bf k}_F^n$, is plotted as a function
of ${\bf k}$ from $\Gamma$ to ${\bf k}_F^n$. 
Clearly, the background increases 
from ${\bf k}_F^n$ to $\Gamma$.
Other unknown effects\cite{add}, contributing to the background, are 
probably present because  the predicted quasiparticle intensity 
near $\Gamma$ seems to be larger 
than in the experiment however, the calculation shows, qualitatively, 
common features with the experiment. 
The preceding discussion and the existence of the quasiparticle near $\Gamma$ put  
our results closer to the interpretation given by Pan {\it et al.}; near 
the Fermi surface a strongly renormalized parabolic coherent band is present 
and the vertical ``dive'' is likely
the incoherent part.

\begin{figure}
\vspace{1cm}
\begin{center}
\setlength{\unitlength}{1cm}
\includegraphics[width=8cm,angle=0]{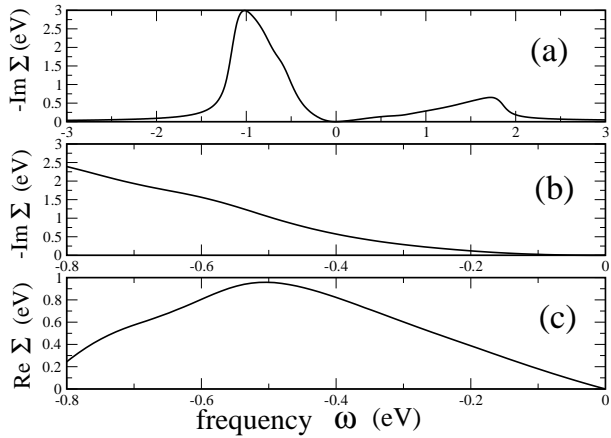}
\end{center}
\caption{
(a) $-Im \Sigma(k_F^n,\omega)$ vs $\omega$ in the full range of frequency.
The self-energy is strongly asymmetric around $\omega=0$ reflecting differences 
between  
the addition and removal of a single electron in a correlated system. 
$Im \Sigma$ shows large structures at large energies and no 
sign of saturation up to energies of 
$ \sim - 1eV$ is observed. These large structures are the responsible 
for the incoherent features described in Fig.\ref{SF} and are mainly due to  
collective charge fluctuations (see text for discussions). 
Since high energy features are due to collective charge fluctuations 
they are very robust against the value of $J$. 
(b) and (c) are $-Im \Sigma(k_F^n,\omega)$,   
and $Re \Sigma(k_F^n,\omega)$ vs $\omega$ respectively,  
for $-0.8eV < \omega <0$.  
}\label{SE}
\end{figure}

In recent $ARPES$ experiments \cite{meevasana} high energy features were 
discussed 
as a function of doping
showing that their energy position decreases
with increasing doping 
(see Fig.1 in that paper).  
This behavior is consistent with the 
expected one in our calculation (see Figs.3-5 in Ref.[\onlinecite{bejas06}]). 
On the other hand, 
in Ref.[\onlinecite{meevasana}], it  was also obtained that high energy 
features lie 
at higher energies than the 
predictions of the band structure calculations. 
This behavior, which is anomalous 
because interactions 
should reduce the bandwidth, to our opinion,
may be considered as an additional support for the interpretation of the 
high energy features 
in terms of incoherent structures due to electronic correlations. 

Self energy results are presented in Fig.\ref{SE}
for ${\bf k}={\bf k}_F^n$.  
In panel (a), $-Im \Sigma({\bf k}_F^n,\omega)$, in the full range of frequency, is 
shown. $\Sigma({\bf k},\omega)$ is strongly asymmetric with respect to $\omega=0$
which is due to the difference between the addition and 
removal of a single electron in a correlated system. 
Notice that the self-energy presents large structure at large energy
with no sign  of saturation up to energies of $\sim -1eV$.
For a better comparison with the self-energy behavior reported by the recent 
$ARPES$ measurements, the imaginary and real parts  of the self energy are presented 
in panels (b) and (c), respectively, for the energy range $-0.8 eV <\omega <0$.
Interestingly, both $Re \Sigma$ and $Im \Sigma$ 
present 
similar shape and order of magnitude to the experiment\cite{valla,xie}.
For instance,  $Re \Sigma$ (Fig.\ref{SE}c) shows a maximum 
of the order of $\sim 0.9 eV$
at about $\sim -0.5 eV$. (In Fig.\ref{exp}c, Figs.4b and 4c from 
Ref.[\onlinecite{valla}] are reproduced for comparison with Fig.\ref{SE}). 
This behavior, as discussed in Ref.[\onlinecite{valla}], 
is  
in contrast with previous reports were 
the self-energy spread out over a much  lower energy scale\cite{kordyuk05}. 
Therefore, our self-energy shows a large energy scale of the order of $1 eV$ 
which is 
responsible for the incoherent structure showed by the spectral functions.

Finally, at low energies, the $Re \Sigma$ (Fig.\ref{SE}c) presents only one slope
while in the experiments 
\cite{valla,xie} two slopes can be seen; above and  below $\sim 50 meV$ 
(Fig.\ref{exp}c). 
Since the lower slope is larger than the upper one
\cite{valla,xie},  
we may associate the  upper slope as 
originated by electronic high energy contributions while, at low energy 
there are additional contributions, associated with the former kink 
\cite{valla99,
bogdanov00,
kaminski00}.
According to Ref.[\onlinecite{valla}] 
the spectral weight of these low energy excitations is only $\sim 10\%$ of the full
spectra.
From Fig.\ref{SE}c, our estimated slope is 
$\lambda= -\frac{ \partial Re \Sigma}{\partial \omega} \sim 1.5$ which is close 
to the experimental upper slope  
seen for $\omega >50 meV$. 
We take this fact as an additional support for considering that high energy
features are contained in our description. 
Even when in this paper we are mainly interested on the high energy 
features a few statements about the low energy kink are noteworthy. 
Our self-energy (Fig.\ref{SE}) does not show any low energy scale
however, this is not crucial because, after much discussion, 
the origin of the kink remains open 
\cite{kordyuk05} and one possibility is that 
the kink is due to phonons \cite{shen,zeyher}. 
If this is the case, a pure $tt'-J$ model calculation, as in  present case, 
does not account for the expected low energy self-energy renormalizations. 
From the pure $t-J$ model low energy spectral features  of magnetic 
origin may also be expected\cite{moreo} however, they should be weaker for 
highly overdoped than for underdoped samples which is not clear from the experiments.
According to results in Ref.[\onlinecite{moreo}] the spectral weight of low energy 
spectral features is a fraction of the quasiparticle weight and then, 
presumably small. 

Let us discuss the origin of the high energy features.
In Ref.[\onlinecite{valla}], 
they are associated with magnetic excitations.
However, since these features are present in highly overdoped samples  
where magnetism is very weak, we take this interpretation with 
caution. 
Our model calculation suggests that they are due to charge fluctuations. 
In the usual many body language, self-energy can be expressed in terms of 
$\alpha^2 F(\omega)$, where the notation is such that $F(\omega)$ gives information of
the density of state of a boson interacting with electrons, and $\alpha^2$ about the 
coupling. 
In Ref.[\onlinecite{bejas06}] 
it was shown that collective charge fluctuations, playing the role of 
bosonic excitations,
are the main contribution to 
$\alpha^2 F(\omega)$ and they lead to strong incoherent features at high energies.
Collective charge fluctuations discussed here are similar to those reported 
by inelastic x-rays scattering\cite{hasan00,markiewicz}.
Two recent theoretical papers\cite{macridin,markiewicz1} have considered
spin fluctuations for explaining the high energy features. In both calculations, 
self-energy renormalizations suggest an abrupt evolution (high energy kink) 
of the low energy quasiparticle peak to the high energy features. This scenario
is different to the ours, where a low energy coherent quasiparticle band 
coexists with high 
energy incoherent structures.
In addition, based on RPA calculations, charge fluctuations 
were ruled out 
in Ref.[\onlinecite{markiewicz1}]. In contrast, in our approach, collective charge 
fluctuations, when they are treated in the strong coupling limit of the $t-J$ model, 
lead to a self-energy (more stronger and asymmetric than RPA) which
produces the results discussed in present paper.

It is important to test, to what extent, $\Sigma({\bf k},\omega)$ from $ARPES$ is 
compatible with transport measurements. 
In order to get some insight into this problem,
the resistivity $\rho$ vs temperature $T$ is shown in Fig.\ref{SFp}c.
For estimating the resistivity it was used the  
expression
$\rho(T)=\frac{4\pi}{\omega_p^2} \frac{1}{\tau}(T)$, 
where 
$\frac{1}{\tau}(T)=-2 Im \Sigma(\omega=0,T)$.
It is known that $1/\tau$ is related with the $Im (\Sigma)$ averaged over the 
Fermi surface 
using the weight factor $(1-cos \theta)$, however 
our self-energy 
is very isotropic over the Fermi surface and then,  $1/\tau$ is very close to  
$-2Im(\Sigma)$. 
For the plasma frequency we choose $\omega_p=2 eV$
\cite{Hwang,Ma}. 
In spite of the approximations and the fact that impurities may also contribute, 
the resistivity values    
are in the order of magnitude of  
the experiments\cite{Ma,Timusk}.
Interestingly, we found a fractional power law behavior $\rho \sim T^{m}$ with $m=1.6-1.7$ 
which is close to that reported in Refs.[\onlinecite{yang06,Ma}] for  
overdoped regime. This fractional power law was discussed in Ref.[\onlinecite{castro04}] 
as indication 
of an anomalous Fermi liquid behavior in overdoped cuprates.

In conclusion, 
we have studied high energy $ARPES$ spectral features in the framework of the 
$tt'-J$ model. Results for spectral functions $A({\bf k},\omega)$ and self-energy 
$\Sigma({\bf k},\omega)$ were presented and confronted with the experiments.  
A strongly renormalized quasiparticle parabolic band was obtained near the 
Fermi surface and 
incoherent structures exist at large energy ($\sim -1eV$). 
The present results support the experimental interpretation given by  
Pan {\it et al.}\cite{pan}.

The author thanks to M. Bejas and A. Foussats for valuable discussion and H. Parent
for critical reading the manuscript.

\end{document}